\documentclass[reprint,aps,prd,floats,floatfix,amsmath,amssymb,amsfonts,nofootinbib,longbibliography,superscriptaddress]{revtex4-1}

\usepackage{graphicx}
\usepackage{dcolumn}
\usepackage{bm}
\usepackage{tensor} 
\usepackage{natbib}
\usepackage{subcaption}

\usepackage[hyperindex,colorlinks]{hyperref}

\begin{document}

\title{Stellar structure in a Newtonian theory with variable $G$}

\author{J\'ulio C. Fabris}
\email{julio.fabris@cosmo-ufes.org}%
\affiliation{%
N\'ucleo Cosmo-ufes \& Departamento de F\'isica,  Universidade Federal do Esp\'irito Santo (UFES)\\
Av. Fernando Ferrari, 540, CEP 29.075-910, Vit\'oria, ES, Brazil.}%
\affiliation{%
National Research Nuclear University MEPhI, Kashirskoe sh. 31, Moscow 115409, Russia}%
\author{Túlio Ottoni}%
\email{tulio.costa@edu.ufes.br}
\affiliation{%
PPGCosmo, CCE, Universidade Federal do Esp\'irito Santo (UFES)\\
Av. Fernando Ferrari, 540, CEP 29.075-910, Vit\'oria, ES, Brazil.}%

\author{Júnior D. Toniato}
\email{junior.toniato@ufop.edu.br}
\author{Hermano Velten}%
\email{hermano.velten@ufop.edu.br}
\affiliation{%
Departamento de F\'isica, Universidade Federal de Ouro Preto (UFOP), Campus Universit\'ario Morro do Cruzeiro, 35.400-000, Ouro Preto, Brazil}%
\date{\today}

\begin{abstract}

A Newtonian-like theory inspired by the Brans-Dicke gravitational Lagrangian has been recently proposed in Ref. \cite{jjtv}. For static configurations the gravitational coupling acquires an intrinsic spatial dependence within the matter distribution. Therefore, the interior of astrophysical configurations may provide a testable environment for this approach as long as no screening mechanism is evoked.
In this work we focus on the stellar hydrostatic equilibrium structure in such varying $G$ scenario. A modified Lane-Emden equation is presented and its solutions for various values of the polytropic index are discussed. The role played by the theory parameter $\omega$, the analogue of the Brans-Dicke parameter, on the physical properties of stars is discussed.   
\end{abstract}

\maketitle

\section{Introduction}

Are the fundamental constants of physics really constants? This is a quite old question, perhaps as older as the identification 
of those constants themselves. In Physics, we can identify, in special, four fundamental constants, each one connected with a given theoretical structure: $\hbar$, which defines the quantum world; $c$, the velocity of light, that is the limit velocity and is related to the relativistic domain; $G$ which indicates the presence of the gravitational interaction; $k_B$, the Boltzmann constant related to
thermodynamics. 
The presence of one or more of these constants in a given equation can suggest what sort of phenomena we are dealing with. For example, an equation containing $G$ refers to gravitation. If besides it contains $c$, we are facing a relativistic gravitational structure, like general relativity. If we add $\hbar$, we are discussing quantum gravity. A phenomena which, by its nature is relativistic and involves gravitation and quantum mechanics and moreover has a thermodynamic characteristics will contain those four constants. This is the case, for example, of Hawking radiation in a black hole.

Among these four constants, the gravitational coupling $G$ was the first one to be identified, although it is the one which is known with the poorest precision: its value is determined up to the $10^{-4}$ order \cite{10.1093/nsr/nwaa165}. This is a consequence of the universality of this fundamental physical interaction, the only one that is rigorously present in all phenomena in nature, and always with an attractive behavior. These features led to identify gravity with the innermost nature of space and time: all modern theories interpret the gravitational phenomena as a consequence of  spacetime curvature.
Moreover, since it is always attractive, it dominates the behavior of large scale systems, like in astrophysics and cosmology.

There are very stringent observational and experimental constraints on the variation of $G$. In spite of these constraints, even a small variation of $G$ with time and/or position may have significant impact on the cosmological and astrophysical observables. For example, the $H_0$ tension may be highly suppress if $G$ varies with time \cite{PhysRevD.104.L021303}. Even the structure formation scenario may change substantially if $G$ is not a constant. There are many relativistic theories of gravity which try to incorporate the variation of the gravitational coupling. The traditional paradigm of such theoretical formulation is the Brans-Dicke theory \cite{Brans:1961sx}, based on a original proposal made by Dirac, inspired by the large number hypothesis which signs out some curious coincidence of numbers obtained from the combination of the constants and some functions of time evaluated today, like the Hubble constant \cite{Dirac:1937ti, Dirac:1938mt}. The Brans-Dicke theory is a more complete formulation of theoretical developments made by Jordan using the idea that $G$ may not be a constant. Today, the Horndesky class of theories \cite{Horndeski:1974wa} provides the most general gravitational Lagrangian leading to second order differential field equations. In most of the cases,
the Horndesky theories incorporate the possibility of a dynamical gravitational coupling.

Even if there exists such plethora of relativistic theories with varying gravitational coupling, it is not so easy to construct a Newtonian theory with a dynamical $G$. To our knowledge, the first proposals to incorporate a varying $G$ effect in a Newtonian context has been made in Refs. \cite{Landsberg:1975,McVittie:1978,Duval:1990hj}. For example, in Ref. \cite{McVittie:1978} the implementation of this idea was quite simple: in the Poisson equation a constant $G$ is replaced by a varying gravitational coupling, a function $G(t)$. There is no dynamical equation for this new function, whose behavior with time must be imposed ad hoc. A natural choice is to use the Dirac proposal, with 
\begin{eqnarray}
G = G_0\,\frac{t_0}{t},
\end{eqnarray}
where $t_0$ is the present age of the universe and $G_0$ is the present value of the gravitational coupling.
This Newtonian theory with varying gravitation coupling has no complete Lagrangian formulation, since $G(t)$ is an arbitrary function.

In a recent paper \cite{jjtv}, a new Newtonian theory with a varying gravitational coupling has been proposed. The gravitational coupling, now given by a function of time and position, is dynamically determined together with the gravitational potential from a new gravitational Lagrangian. It has been shown that this theory is consistent with the general properties of spherical objects like stars and, at same time, its homogeneous and isotropic cosmological solutions can generate an accelerated expansion of the universe.

Of course, one may wonder about the interest to construct a Newtonian theory with varying $G$. One can evoke the academic interest of obtaining a complete and consistent Newtonian formulation implementing a dynamical gravitational coupling. The Newtonian framework is, in principle, simpler than the relativistic one, so why it would be so difficult to give a dynamical behaviour to $G$, something that is, if not trivial, at least perfectly possible in a relativistic context? But we can quote at least two other motivations. First,
a consistent Newtonian theory with varying $G$ may suggest possible new relativistic structures as, for example, the non-minimal coupling of gravity with other fields, in a similar way as somehow the Poisson equation suggested the general relativity equations. Second, many astrophysical and cosmological problems are more conveniently analyzed in a Newtonian framework, e.g., the dynamics of galaxies, clusters of galaxies and even numerical simulations of the large scale structure. If $G$ is not a constant, it would be nice
to have a consistent Newtonian theory incorporating this feature.

In this paper, we focus on the stellar structure of non-relativistic stars. This is an important analysis in the context of the theory proposed in \cite{jjtv} since it has been shown that the main difference from the standard Newtonian gravity should manifest within matter distributions.

\section{Newtonian theory with variable \textit{G}}\label{sec:classical_g}

In Ref.~\cite{jjtv} a Lagrangian for a theory with varying $G$ has been proposed and we review it in this section.
The Lagrangian of this new approach is given by
\begin{equation}\label{LagrangianG}
{\cal L} = \frac{\nabla\psi\cdot\nabla\psi}{2} - \frac{\omega}{2}\biggr(\psi\frac{\dot\sigma^2}{\sigma^2} - c^4\nabla\sigma\cdot\nabla\sigma \biggl) + \ 4\pi G_0 \rho \sigma \psi,
\end{equation}
where $G_0$ is a constant, $\psi$ is an equivalent of the ordinary Newtonian potential and $\sigma$ is a new function related to the gravitational coupling. Also, we have introduced the parameter $\omega$ which shall be assumed to be constant.

At the Lagrangian level, it is clear that usual Newtonian gravity is recovered with the identifications $\omega=0$ and $\sigma=1$. However, as already discussed in Ref. \cite{jjtv} and as we will also verify later, the standard Newtonian limit takes place with $\sigma$ constant and $\omega \rightarrow \infty$. A constant with dimensions of velocity, the speed of light $c$, has been introduced to guarantee the Lagrangian has the correct physical dimensions. But, no direct mention is made to a relativistic framework in doing so: one has just borrow from electromagnetism two fundamental constants, the vacuum electric permissivity $\epsilon_0$ and magnetic permeability $\mu_0$.

In some sense the Lagrangian above corresponds to the Newtonian version of the relativistic Brans-Dicke theory (in Einstein's frame). We emphasize again that the constant $c$ appears in this Lagrangian for dimensional reasons. This does not mean this is a relativistic theory since this Lagrangian is invariant under the Galilean group transformations.

Applying the Euler-Lagrange equations of motion, 
\begin{align}
\nabla\cdot\frac{\partial{\cal L}}{\partial \nabla\psi} - \frac{\partial{\cal L}}{\partial\psi} &= 0,\\[1ex]
\frac{d}{dt}\frac{\partial{\cal L}}{\partial\dot\sigma} + \nabla\cdot\frac{\partial{\cal L}}{\partial \nabla\sigma} - \frac{\partial{\cal L}}{\partial\sigma} &= 0,
\end{align}
the following equations are obtained,
\begin{align}
\label{en1}
\nabla^2\psi + \frac{\omega}{2}\left(\frac{\dot\sigma}{\sigma}\right)^2  &= 4\pi G_0 \sigma \rho,\\[1ex]
\label{en2}
c^4\frac{\sigma}{\psi}\nabla^2\sigma - \frac{d}{dt}\biggr(\frac{\dot\sigma}{\sigma}\biggl) - \frac{\dot\psi}{\psi}\frac{\dot\sigma}{\sigma} &= \frac{4\pi G_0\sigma \rho}{\omega}.
\end{align}
The over-dot indicate total time derivative, which assures to the resulting equations an invariance with respect to Galilean transformations. Equations \eqref{en1} and \eqref{en2} show that the quantity $G_0\sigma$ can be interpreted as an effective gravitational coupling.

It is worth noting that the above set of equations can not be seen as the non-relativistic limit of a pure covariant scalar-tensor gravitational theory. Due to the limiting cases for $\sigma$ and $\omega$ in obtaining the Newtonian behavior as it happens with the Brans-Dicke theory, only a self-similarity with Brans-Dicke can be evoked. The true scalar-tensor theory giving rise to this non-relativistic dynamics is still missing.

\section{Gravitational field within mass distributions}

As already pointed out in Ref. \cite{jjtv}, equations \eqref{en1} and \eqref{en2} are decoupled when in vacuum. Thus, the fields $\psi$ and $\sigma$ have independent dynamics with both satisfying Laplace's equations. Only within matter their dynamics are linked. One should therefore focus on such interior solutions to better understand the behaviour of the theory.

\subsection{Constant density sphere}


Let us start by reviewing the simple realization of a static sphere of radius $R$ with constant density $\rho_0$. As shown in \cite{jjtv} the gravitational potential in this case assumes the form

\begin{align}
    \psi(r) =& -\frac{G_0M\,k}{kR\cosh{(kR)}-\sinh{(kR)}}\times  \nonumber \\[1ex] & \ \left[\cosh(kR)-\frac{\sinh{(k\,r)}}{k\,r}\right],\quad \mbox{for}\quad r<R,\label{psi-int}\\[2ex]
    \psi(r) =& -\frac{G_0M}{r},\quad \mbox{for}\quad r\geq R.
\end{align}
In the above result we have defined
\begin{eqnarray}
k^2 = \frac{4\pi G_0\rho_0}{c^2\sqrt{\omega}},
\end{eqnarray}
which is valid for the case $\omega>0$. 

By approximating the quantity $kr$ such that
\begin{equation}
    kr\sim \frac{10^{-2}}{\omega^{1/4}}\sqrt{\frac{M/M_\odot}{R/R_\odot}}\,\left(\frac{r}{R}\right),
\end{equation}
where $M_\odot$ and $R_\odot$ are the mass and radius of the Sun, an order of magnitude estimation for possible deviations from the standard Newtonian gravity can be found. For constant density star configurations this occurs for,
\begin{equation}\label{MtoR}
    \frac{M}{R}\gtrsim 10^4\sqrt{\omega}\frac{M_\odot}{R_\odot}.
\end{equation}

According to this rough estimation, \eqref{MtoR} means that the Sun can not probe manifestations of this theory if the theory parameter assume values $\omega \gtrsim 10^{-8}$. Only more compact objects would be suitable for testing the theory. However, more realistic scenarios should be investigated. This is the goal of the next sections.

\subsection{The modified Lane-Emden equation}
In ordinary Newtonian gravity, the Lane-Emden equation is a useful description of self-gravitating spheres. It is constructed by assuming a polytropic fluid as a source of the gravitational potential, where pressure and density are related through the expression
\begin{equation}
    p=K\rho^{1+1/n},\label{polytropic}
\end{equation}
with $K$ a constant and $n$ the so called polytropic index.
Moreover, the Lane-Emden equation is dimensionless, which is a suitable property for numerical analysis. Thus, in this section, we will show how the usual Lane-Emden equation is modified in the varying-$G$ Newtonian gravity. 

Assuming a static and spherically symmetric star, all functions depend only on the radial coordinate $r$. The momentum conservation (Euler's equation) for this distribution assumes a static velocity field with $\vec v = 0$, leading to the relation
\begin{eqnarray}
\label{em-s1}
\frac{p'}{\rho} &= - \psi',
\end{eqnarray}
where the symbol prime $(\, ^{\prime} \,)$ means a derivative with respect to $r$. It is worth noting we are assuming that only the gradient of the potential $\psi$ is relevant for the classical hydrostatic equilibrium. 
One should therefore obtain the behavior of the potential $\psi$ which is coupled to the field $\sigma$. This can be derived from \eqref{en1} and \eqref{en2}, which can be rewritten as a new set of equations. They read
\begin{eqnarray}
\label{em-s2}
\psi'' + 2\frac{\psi'}{r}  &= 4\pi G_0 \sigma \rho,\\
\label{em-s3}
\sigma'' + 2\frac{\sigma'}{r}  &= \frac{4\pi G_0}{c^4\omega}\psi \rho.
\end{eqnarray}
As mentioned before, the Newtonian counterpart of the above equations is obtained when $\sigma$ is a constant and $\omega$ tends to infinity.

To proceed, we redefine the density as
\begin{eqnarray}
\rho(r) = \rho_c\Theta (r)^n,\label{rho-theta}
\end{eqnarray}
where $\Theta(r)$ is a dimensionless density function such that $\Theta(0)=1$. Thus, $\rho_c$ represents the central density value.
With this redefinition and the polytropic equation of state \eqref{polytropic}, Euler equation \eqref{em-s1} can be integrated, resulting in a relation between $\psi$ and $\Theta$, 
\begin{equation}
    \psi(r)=-K  (n+1)\rho_c^{1/n}\Theta(r) +\psi_R.
\end{equation}

The parameter $\psi_R$ is a constant of integration that must be fixed by the potential at the star's radius.

With the last result, one can rewrite equations \eqref{em-s2} and \eqref{em-s3} in a similar form to the original Lane-Emden equation,
\begin{eqnarray}
\frac{1}{x^2}\frac{d}{dx}\biggr(x^2\frac{d\Theta}{dx}\biggl) &=& - \sigma\Theta^n,\\[1ex]
\frac{1}{x^2}\frac{d}{dx}\biggr(x^2\frac{d\sigma}{dx}\biggl) &=& - \frac{\kappa_0^2}{\omega}\Theta^{n + 1} + \lambda_0\Theta^n.
\end{eqnarray}
In the above equations we have defined
\begin{eqnarray}
x &=& \frac{r}{r_0}, \quad r_0 = \sqrt{\frac{(n + 1)}{4\pi G_0}\frac{p_c}{\rho_c^2}},\\[1ex]
\kappa_0 &=& \sqrt{\frac{4\pi G_0r_0^2p_c(n+1)}{c^4}}=\frac{K (n+1)\rho_c^{1/n}}{c^2} \\
\lambda_0&=&\frac{p_c(n+1)\psi_R}{\rho_c\omega c^4}
\end{eqnarray}
In order to guarantee that the ordinary Lane-Emden equation is recovered when $\sigma$ is constant and $\omega \rightarrow \infty$, $\lambda_0$ must be set equal to zero. The constant $\kappa_0$ is directly related to the central pressure/central density ratio, namely
\begin{eqnarray}
\frac{p_c}{\rho_c} = \frac{\kappa_0 c^2}{(1 + n)}.
\end{eqnarray}

The dimensionless radius of the star is defined as the point $x_1$ where $\Theta(x_1)=0$. The physical radius of the star is simple
\begin{eqnarray}
R = r_0 x_1.
\end{eqnarray}

An expression for the stellar mass can be obtained by the integral,
\begin{eqnarray}
M = 4\pi\int_0^R\rho(r)r^2dr = 4\pi \rho_cr_0^3\int_0^{x_1}\Theta^n x^2 dx.
\end{eqnarray}

Outside the star, the vacuum external solution for the $\sigma$ field is,
\begin{eqnarray}
\sigma(x) = \sigma_1 + \frac{\sigma_2}{x},
\end{eqnarray}
where $\sigma_1$ and $\sigma_2$ are constants of integration. We will consider $\sigma_1=1$, without loss of generality. However, the external solution must be continuous with the numerical internal solution at the star radius $r=R$. So, we can fix the second constant,
\begin{eqnarray}
\sigma_2 = R(\sigma(R)-1).
\end{eqnarray}
If we now impose the continuity of the derivative of $\sigma$ at the star's radius, we reach a criteria that the numerical solution and its derivative must satisfy
\begin{eqnarray}\label{criteria}
\frac{d\sigma}{dx}\bigg|_{R} = \frac{1-\sigma(R)}{R}.
\end{eqnarray}

Now, for a given central value of $\sigma(0)\equiv\sigma_C$, the solution will give the correct asymptotic behaviour if (\ref{criteria}) is satisfied.

\begin{figure*}[!t]
\begin{subfigure}{.44\textwidth}
\centering
\includegraphics[width=\linewidth]{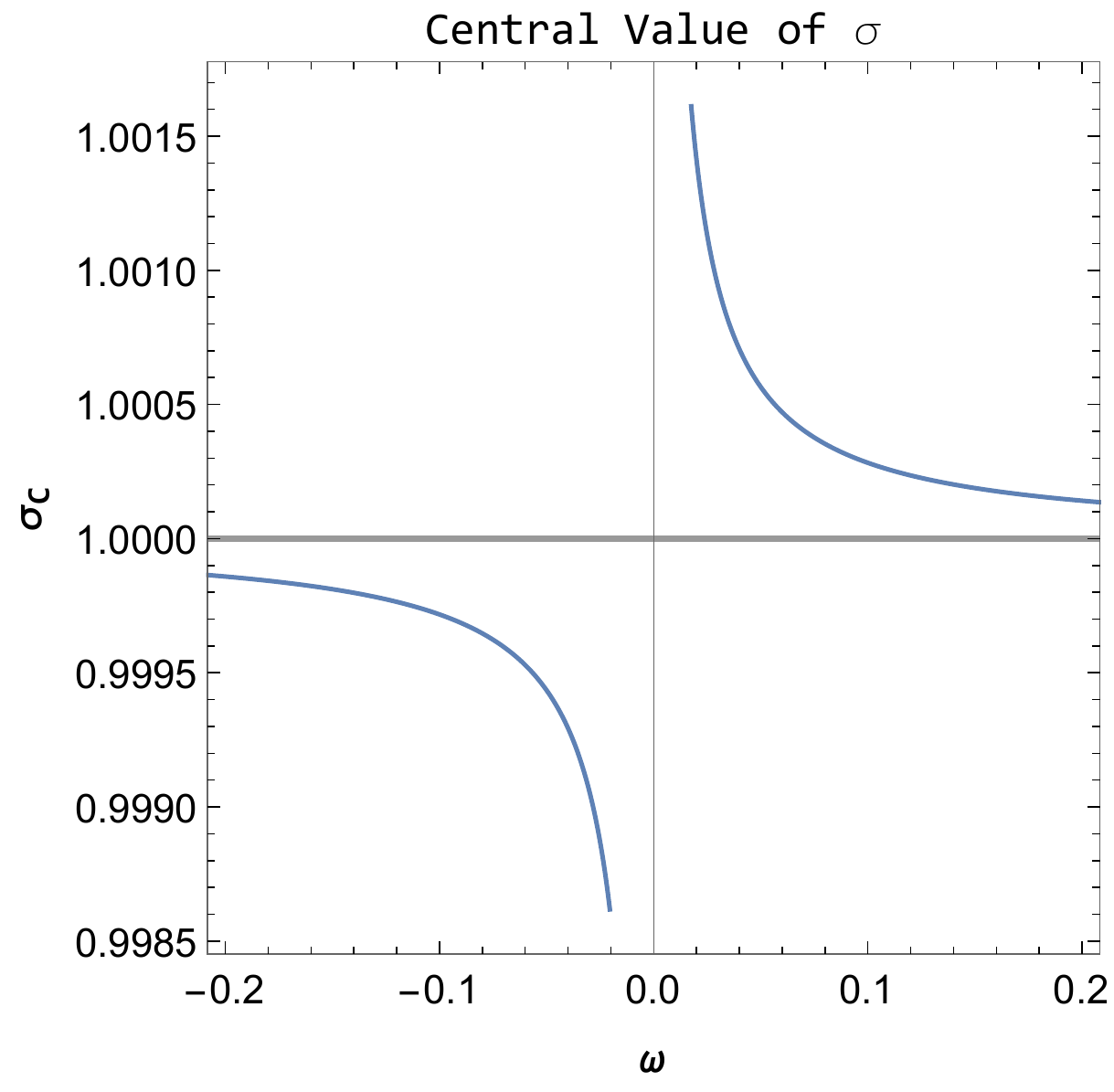}
\end{subfigure}
\hfill
\begin{subfigure}{.44\textwidth}
\centering
\includegraphics[width=\linewidth]{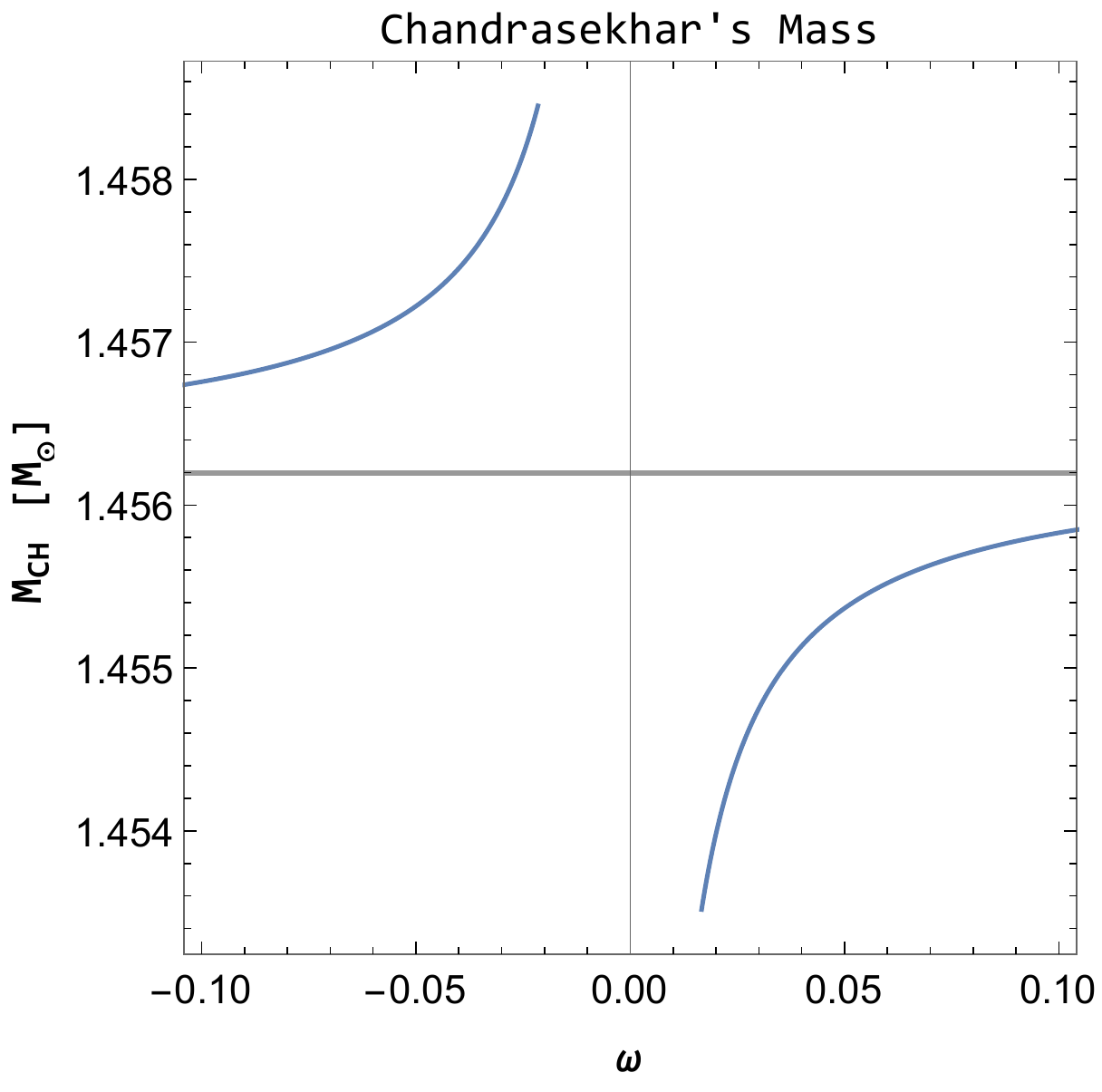}
\end{subfigure}
\caption{Left: The behaviour of the central value of the $\sigma$ field. For positive $\omega$ the gravitational coupling is larger than the Newtonian value and for negative $\omega$ is weaker. Right: The Chandrasekhar mass $M_{CH}$ of such limit star configuration. For positive $\omega$ the mass is smaller than for negative values.}
\label{fig:1}
\end{figure*}

\section{Numerical results}

\subsection{The case $n=1$}

Let us now focus on a specific equation of state with polytropic index $n=1$
\begin{eqnarray}
p = K\rho^2.
\end{eqnarray}
With the adapted Lane-Emden system at hand one can apply them to a couple of typical stellar structures.

Let us firstly consider stars that are similar to the Sun keeping in mind that the chosen equation of state only very crudely can
represent such stars. We are interested in orders of magnitudes estimations.

If the Sun is physically described by the following central density and pressure,
\begin{eqnarray}
&\rho_c \sim 10^{5}\,\mbox{kg/m$^3$},\\
&p_c \sim 10^{16}\,\mbox{Pa},
\end{eqnarray}
after performing a numerical integration, we find that the dimensionless radius of this star is of the order of $x_1 = 0.01$, leading
to $R \sim 10 ^{8}$\,m,  which is roughly the radius of the Sun ($R_\odot = 7\times 10^{8}$\,m).  At the same time the field
$\sigma$ remains essentially constant across the star, which is in agreement with the constraint given by stellar evolution for the Sun. In fact, for the values chosen above, the difference on the value of the gravitational coupling at the center of the star and at infinity is in the fourth decimal case for $|\omega|$ of the order of unity, lesser than the experimental precision obtained so far in the measurements of $G$. The value at the center of the star is approximately $0,01\%$ smaller for $\omega = -1$ and $0,01\%$ greater for
$\omega = 1$, always supposing the standard value at infinity.

If we turn now our attention to even more compact objects we have,
\begin{eqnarray}
&\rho_c \sim 10^{17}\,\mbox{kg/m$^3$},\\
&p_c \sim 10^{34}\, \mbox{Pa}.
\end{eqnarray}
Performing again the numerical integration, we obtain the $x_1 \sim 1$, leading to $R \sim 10$km, roughly in agreement with the expected value. At the same time the field $\sigma$ has a variation along the star of the order of 0.3\%. Hence, compact objects can help to test the theory.

\subsection{The case n=3}

For n=3 we have the equation of state,
\begin{eqnarray}
P=K\rho^{4/3},
\end{eqnarray}
which may represent a completely degenerate and non interacting Fermi gas in the ultra relativistic limit. This is the limit case of white dwarfs, compact stars that sustains itself against gravity by the degeneracy pressure of the electrons. In this case, the constant K have a value of \cite{stellarstr}
\begin{eqnarray}
K=\frac{hc}{8m_u^{4/3}}\left(\frac{3}{\pi}\right)^{1/3} \frac{1}{\mu_e^{4/3}},
\end{eqnarray}
where $m_u$ is the atomic mass unit and $\mu_e$ is the mean molecular weight per free electron. For a completely hydrogen depleted gas $\mu_e \approx 2$.

\begin{figure*}[!t]
\begin{subfigure}{.44\textwidth}
\centering
\includegraphics[width=\linewidth]{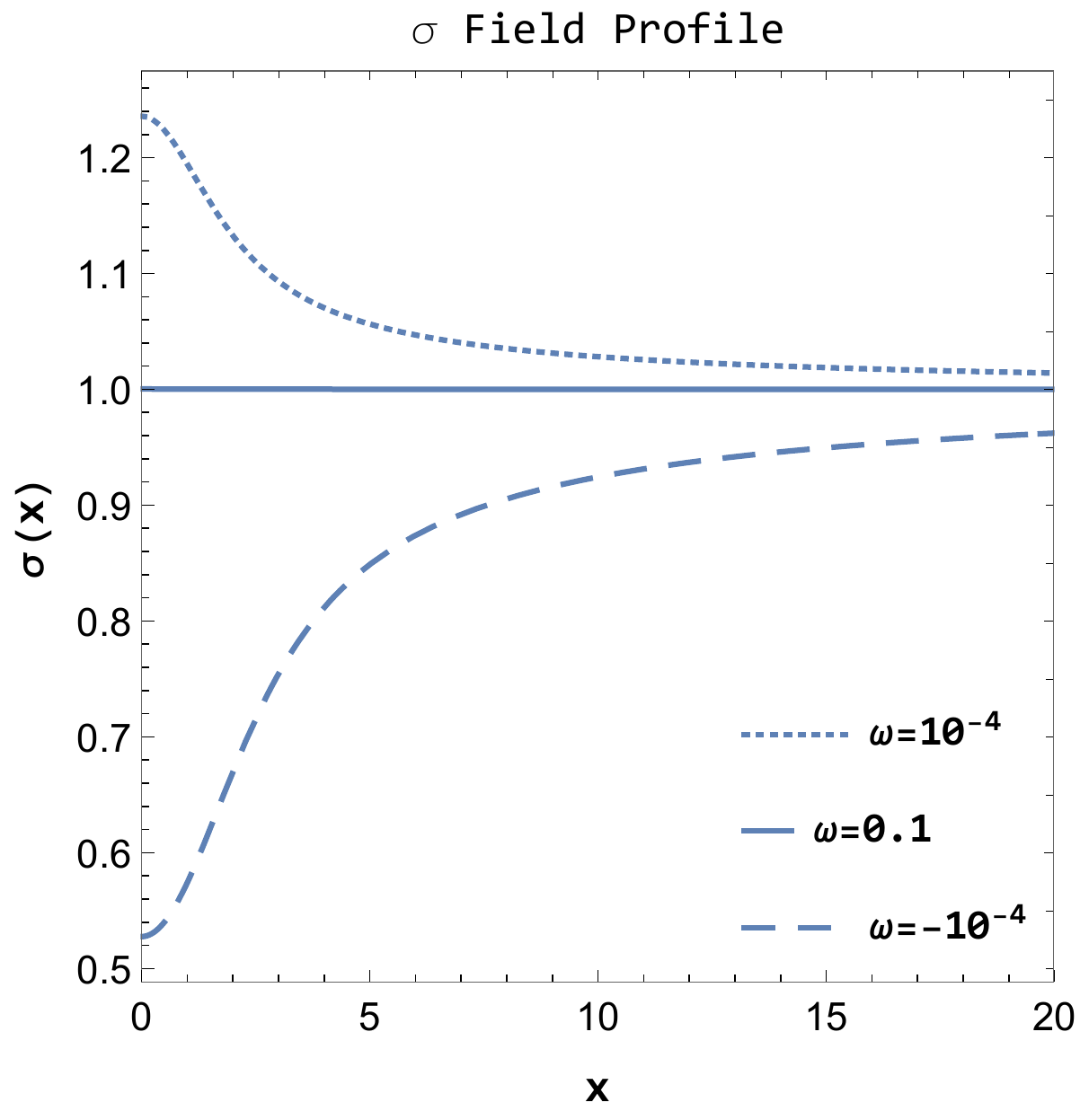}
\end{subfigure}
\hfill
\begin{subfigure}{.44\textwidth}
\centering
\includegraphics[width=\linewidth]{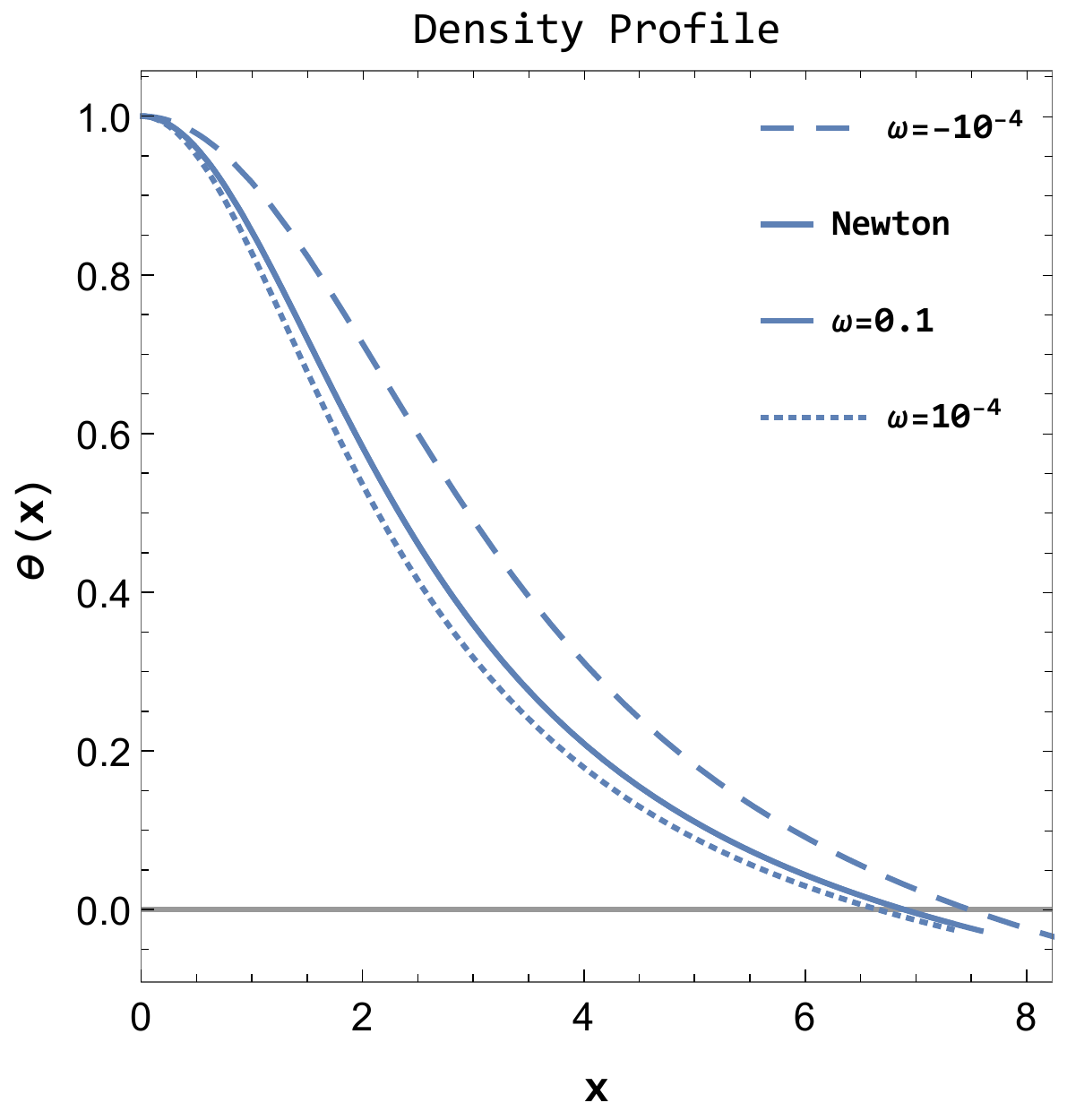}
\end{subfigure}
\caption{Left: The $\sigma$ field profile, with the approximate dimensionless radius of the star marked. For $\omega$ small as $10^{-4}$ the gravitational coupling can be $20\%$ bigger than the asymptotic Newtonian value. For negative small $\omega$, the gravitational coupling at the star's center is almost half the usual value. Right: The dimensionless density profile for the same values of $\omega$. For values of $\omega$ of the order $0.1$ the theory is practically indistinguishable with the Newtonian one.}
\label{fig:2}
\end{figure*}

The results for this case are shown in Fig \ref{fig:1} and \ref{fig:2}. As expected, the theory with varying gravitational coupling is almost identical to the ordinary Newtonian one as the $\omega$ parameter value increases. For small and positive values of $\omega$, the central value of the $\sigma$ field starts to grow and, if it is interpreted as being proportional to the effective gravitational coupling $(G_{eff}=G_0\sigma)$, this indicates a stronger gravity that lowers the star's mass and radius. The situation for negative $\omega$ is the opposite. The central value of $\sigma$ starts to decrease indicating a weaker gravity that increases the star's mass and radius.

Compared results for both positive and negative $\omega$ values are not symmetrical. We have checked that the solutions are more sensitive to negative values of the $\omega$ parameter.

\section{Final remarks}

Even though the actual description of the gravitational phenomena demands a covariant and relativistic formulation the Newtonian gravity still works with acceptable accuracy for a broad range of astrophysical applications. Ref. \cite{jjtv} proposed a non-relativistic version of a modified gravity theory inspired by the Brans-Dicke relativistic scalar-tensor theory of gravity. For simplicity, one can mention two new features of this theory: it possesses a new parameter $\omega$ and the strength of the effective gravitational coupling dictated by the field $\sigma$.

In this work we have applied this non-relativistic theory to the structure of stellar configurations. While in the exterior vacuum solutions both $\psi$ and $\sigma$ satisfy the Laplace's equation, with their behaviour resembling the standard Newtonian potential, deviations are present in the interior solutions. Therefore, one can not probe such new gravitational effects due to the existence of an intrinsic degeneracy with the equations of state for the stellar fluid. On the other hand, by fixing the equation of state it is possible to measure to impact of the theory parameter $\omega$ on the astrophysical observable like star's mass and star's radius.

The manifestation of the new gravitational features depends on the compactness of the star. Curiously such dependence is not present in other modified gravity theories. 

As our main result we have discussed the impact of the $\omega$ parameter on the Chandrasekhar's mass limit. If $\omega=10^{-4}$ we find $M_{CH}=1.14 M_{\odot}$ while for $\omega=10^{-3}$ we find $M_{CH}=1.41 M_{\odot}$   Then, the existence of white dwarfs with masses around $1.3 M_{\odot}$ \cite{10.1111/j.1365-2966.2006.11388.x,Tang_2014} clearly rules out $\omega$ value of order $10^{-4}$ or smaller. On the other hand, higher Chandrasekhar mass limits are allowed for negative $\omega$ values. This case would become very interesting in case of a future detection of a white dwarf more massive than the currently accepted Chandrasekhar limit.

\begin{acknowledgments}
In writing the present article to a special issue in honour of Maxim Khlopov, we have in mind his vast interest in physics ranging from particle physics to cosmology, always having an open attitude for new ideas. We think the ideias exposed in this article they point out to the direction of the search of new physical structures which are suggested by the problems existing in particle physics, astrophysics and cosmology, and in this sense we believe that is adapted to this special issue dedicated to the 70th birthday of Maxim Khlopov. By the way Maxim's pioneer project {\it Virtual Institute of Astroparticle physics} is one of first attempts to organize online seminars, something that became so familiar during this COVID19 pandemic.

The authors thank FAPES/CNPq/CAPES and Proppi/UFOP for financial support. 
\end{acknowledgments}

\bibliography{Refs}

\begin{thebibliography}{13}%
\makeatletter
\providecommand \@ifxundefined [1]{%
 \@ifx{#1\undefined}
}%
\providecommand \@ifnum [1]{%
 \ifnum #1\expandafter \@firstoftwo
 \else \expandafter \@secondoftwo
 \fi
}%
\providecommand \@ifx [1]{%
 \ifx #1\expandafter \@firstoftwo
 \else \expandafter \@secondoftwo
 \fi
}%
\providecommand \natexlab [1]{#1}%
\providecommand \enquote  [1]{``#1''}%
\providecommand \bibnamefont  [1]{#1}%
\providecommand \bibfnamefont [1]{#1}%
\providecommand \citenamefont [1]{#1}%
\providecommand \href@noop [0]{\@secondoftwo}%
\providecommand \href [0]{\begingroup \@sanitize@url \@href}%
\providecommand \@href[1]{\@@startlink{#1}\@@href}%
\providecommand \@@href[1]{\endgroup#1\@@endlink}%
\providecommand \@sanitize@url [0]{\catcode `\\12\catcode `\$12\catcode
  `\&12\catcode `\#12\catcode `\^12\catcode `\_12\catcode `\%12\relax}%
\providecommand \@@startlink[1]{}%
\providecommand \@@endlink[0]{}%
\providecommand \url  [0]{\begingroup\@sanitize@url \@url }%
\providecommand \@url [1]{\endgroup\@href {#1}{\urlprefix }}%
\providecommand \urlprefix  [0]{URL }%
\providecommand \Eprint [0]{\href }%
\providecommand \doibase [0]{http://dx.doi.org/}%
\providecommand \selectlanguage [0]{\@gobble}%
\providecommand \bibinfo  [0]{\@secondoftwo}%
\providecommand \bibfield  [0]{\@secondoftwo}%
\providecommand \translation [1]{[#1]}%
\providecommand \BibitemOpen [0]{}%
\providecommand \bibitemStop [0]{}%
\providecommand \bibitemNoStop [0]{.\EOS\space}%
\providecommand \EOS [0]{\spacefactor3000\relax}%
\providecommand \BibitemShut  [1]{\csname bibitem#1\endcsname}%
\let\auto@bib@innerbib\@empty
\bibitem [{\citenamefont {Fabris}\ \emph {et~al.}(2021)\citenamefont {Fabris},
  \citenamefont {Gomes}, \citenamefont {Toniato},\ and\ \citenamefont
  {Velten}}]{jjtv}%
  \BibitemOpen
  \bibfield  {author} {\bibinfo {author} {\bibfnamefont {J\'ulio~C.}\
  \bibnamefont {Fabris}}, \bibinfo {author} {\bibfnamefont {Tales}\
  \bibnamefont {Gomes}}, \bibinfo {author} {\bibfnamefont {J\'unior~D.}\
  \bibnamefont {Toniato}}, \ and\ \bibinfo {author} {\bibfnamefont {Hermano}\
  \bibnamefont {Velten}},\ }\bibfield  {title} {\enquote {\bibinfo {title}
  {{Newtonian-like gravity with variable $G$}},}\ }\href {\doibase
  10.1140/epjp/s13360-021-01146-z} {\bibfield  {journal} {\bibinfo  {journal}
  {Eur. Phys. J. Plus}\ }\textbf {\bibinfo {volume} {136}},\ \bibinfo {pages}
  {143} (\bibinfo {year} {2021})},\ \Eprint {http://arxiv.org/abs/2009.04434}
  {arXiv:2009.04434 [gr-qc]} \BibitemShut {NoStop}%
\bibitem [{\citenamefont {Xue}\ \emph {et~al.}(2020)\citenamefont {Xue},
  \citenamefont {Liu}, \citenamefont {Li}, \citenamefont {Wu}, \citenamefont
  {Yang}, \citenamefont {Liu}, \citenamefont {Shao}, \citenamefont {Tu},
  \citenamefont {Hu},\ and\ \citenamefont {Luo}}]{10.1093/nsr/nwaa165}%
  \BibitemOpen
  \bibfield  {author} {\bibinfo {author} {\bibfnamefont {Chao}\ \bibnamefont
  {Xue}}, \bibinfo {author} {\bibfnamefont {Jian-Ping}\ \bibnamefont {Liu}},
  \bibinfo {author} {\bibfnamefont {Qing}\ \bibnamefont {Li}}, \bibinfo
  {author} {\bibfnamefont {Jun-Fei}\ \bibnamefont {Wu}}, \bibinfo {author}
  {\bibfnamefont {Shan-Qing}\ \bibnamefont {Yang}}, \bibinfo {author}
  {\bibfnamefont {Qi}~\bibnamefont {Liu}}, \bibinfo {author} {\bibfnamefont
  {Cheng-Gang}\ \bibnamefont {Shao}}, \bibinfo {author} {\bibfnamefont
  {Liang-Cheng}\ \bibnamefont {Tu}}, \bibinfo {author} {\bibfnamefont
  {Zhong-Kun}\ \bibnamefont {Hu}}, \ and\ \bibinfo {author} {\bibfnamefont
  {Jun}\ \bibnamefont {Luo}},\ }\bibfield  {title} {\enquote {\bibinfo {title}
  {{Precision measurement of the Newtonian gravitational constant}},}\ }\href
  {\doibase 10.1093/nsr/nwaa165} {\bibfield  {journal} {\bibinfo  {journal}
  {National Science Review}\ }\textbf {\bibinfo {volume} {7}},\ \bibinfo
  {pages} {1803--1817} (\bibinfo {year} {2020})},\ \Eprint
  {http://arxiv.org/abs/https://academic.oup.com/nsr/article-pdf/7/12/1803/38880653/nwaa165.pdf}
  {https://academic.oup.com/nsr/article-pdf/7/12/1803/38880653/nwaa165.pdf}
  \BibitemShut {NoStop}%
\bibitem [{\citenamefont {Marra}\ and\ \citenamefont
  {Perivolaropoulos}(2021)}]{PhysRevD.104.L021303}%
  \BibitemOpen
  \bibfield  {author} {\bibinfo {author} {\bibfnamefont {Valerio}\ \bibnamefont
  {Marra}}\ and\ \bibinfo {author} {\bibfnamefont {Leandros}\ \bibnamefont
  {Perivolaropoulos}},\ }\bibfield  {title} {\enquote {\bibinfo {title} {Rapid
  transition of ${G}_{\mathrm{eff}}$ at ${z}_{t}\ensuremath{\simeq}0.01$ as a
  possible solution of the hubble and growth tensions},}\ }\href {\doibase
  10.1103/PhysRevD.104.L021303} {\bibfield  {journal} {\bibinfo  {journal}
  {Phys. Rev. D}\ }\textbf {\bibinfo {volume} {104}},\ \bibinfo {pages}
  {L021303} (\bibinfo {year} {2021})}\BibitemShut {NoStop}%
\bibitem [{\citenamefont {Brans}\ and\ \citenamefont
  {Dicke}(1961)}]{Brans:1961sx}%
  \BibitemOpen
  \bibfield  {author} {\bibinfo {author} {\bibfnamefont {C.}~\bibnamefont
  {Brans}}\ and\ \bibinfo {author} {\bibfnamefont {R.H.}\ \bibnamefont
  {Dicke}},\ }\bibfield  {title} {\enquote {\bibinfo {title} {{Mach's principle
  and a relativistic theory of gravitation}},}\ }\href {\doibase
  10.1103/PhysRev.124.925} {\bibfield  {journal} {\bibinfo  {journal} {Phys.
  Rev.}\ }\textbf {\bibinfo {volume} {124}},\ \bibinfo {pages} {925--935}
  (\bibinfo {year} {1961})}\BibitemShut {NoStop}%
\bibitem [{\citenamefont {Dirac}(1937)}]{Dirac:1937ti}%
  \BibitemOpen
  \bibfield  {author} {\bibinfo {author} {\bibfnamefont {Paul~A.M.}\
  \bibnamefont {Dirac}},\ }\bibfield  {title} {\enquote {\bibinfo {title} {{The
  Cosmological constants}},}\ }\href {\doibase 10.1038/139323a0} {\bibfield
  {journal} {\bibinfo  {journal} {Nature}\ }\textbf {\bibinfo {volume} {139}},\
  \bibinfo {pages} {323} (\bibinfo {year} {1937})}\BibitemShut {NoStop}%
\bibitem [{\citenamefont {Dirac}(1938)}]{Dirac:1938mt}%
  \BibitemOpen
  \bibfield  {author} {\bibinfo {author} {\bibfnamefont {Paul~A.M.}\
  \bibnamefont {Dirac}},\ }\bibfield  {title} {\enquote {\bibinfo {title} {{New
  basis for cosmology}},}\ }\href {\doibase 10.1098/rspa.1938.0053} {\bibfield
  {journal} {\bibinfo  {journal} {Proc. Roy. Soc. Lond. A}\ }\textbf {\bibinfo
  {volume} {A165}},\ \bibinfo {pages} {199--208} (\bibinfo {year}
  {1938})}\BibitemShut {NoStop}%
\bibitem [{\citenamefont {Horndeski}(1974)}]{Horndeski:1974wa}%
  \BibitemOpen
  \bibfield  {author} {\bibinfo {author} {\bibfnamefont {Gregory~Walter}\
  \bibnamefont {Horndeski}},\ }\bibfield  {title} {\enquote {\bibinfo {title}
  {{Second-order scalar-tensor field equations in a four-dimensional space}},}\
  }\href {\doibase 10.1007/BF01807638} {\bibfield  {journal} {\bibinfo
  {journal} {Int. J. Theor. Phys.}\ }\textbf {\bibinfo {volume} {10}},\
  \bibinfo {pages} {363--384} (\bibinfo {year} {1974})}\BibitemShut {NoStop}%
\bibitem [{\citenamefont {Landsberg}\ and\ \citenamefont
  {Bishop}(1975)}]{Landsberg:1975}%
  \BibitemOpen
  \bibfield  {author} {\bibinfo {author} {\bibfnamefont {P.~T.}\ \bibnamefont
  {Landsberg}}\ and\ \bibinfo {author} {\bibfnamefont {N.~T.}\ \bibnamefont
  {Bishop}},\ }\bibfield  {title} {\enquote {\bibinfo {title} {{A Principle of
  Impotence Allowing for Newtonian Cosmologies with a time-Dependent
  Gravitational Constant}},}\ }\href {\doibase 10.1093/mnras/171.2.279}
  {\bibfield  {journal} {\bibinfo  {journal} {Mon. Not. R. astr. Soc.}\
  }\textbf {\bibinfo {volume} {171}},\ \bibinfo {pages} {279--286} (\bibinfo
  {year} {1975})}\BibitemShut {NoStop}%
\bibitem [{\citenamefont {McVittie}(1978)}]{McVittie:1978}%
  \BibitemOpen
  \bibfield  {author} {\bibinfo {author} {\bibfnamefont {G.~C.}\ \bibnamefont
  {McVittie}},\ }\bibfield  {title} {\enquote {\bibinfo {title} {{Newtonian
  cosmology with a time-varying constant of gravitation}},}\ }\href {\doibase
  10.1093/mnras/183.4.749} {\bibfield  {journal} {\bibinfo  {journal} {Mon.
  Not. R. astr. Soc.}\ }\textbf {\bibinfo {volume} {183}},\ \bibinfo {pages}
  {749--764} (\bibinfo {year} {1978})}\BibitemShut {NoStop}%
\bibitem [{\citenamefont {Duval}\ \emph {et~al.}(1991)\citenamefont {Duval},
  \citenamefont {Gibbons},\ and\ \citenamefont {Horvathy}}]{Duval:1990hj}%
  \BibitemOpen
  \bibfield  {author} {\bibinfo {author} {\bibfnamefont {C.}~\bibnamefont
  {Duval}}, \bibinfo {author} {\bibfnamefont {Gary~W.}\ \bibnamefont
  {Gibbons}}, \ and\ \bibinfo {author} {\bibfnamefont {P.}~\bibnamefont
  {Horvathy}},\ }\bibfield  {title} {\enquote {\bibinfo {title} {{Celestial
  mechanics, conformal structures and gravitational waves}},}\ }\href {\doibase
  10.1103/PhysRevD.43.3907} {\bibfield  {journal} {\bibinfo  {journal} {Phys.
  Rev. D}\ }\textbf {\bibinfo {volume} {43}},\ \bibinfo {pages} {3907--3922}
  (\bibinfo {year} {1991})},\ \Eprint {http://arxiv.org/abs/hep-th/0512188}
  {arXiv:hep-th/0512188} \BibitemShut {NoStop}%
\bibitem [{\citenamefont {Kippenhahn}\ \emph {et~al.}(2012)\citenamefont
  {Kippenhahn}, \citenamefont {Weigert},\ and\ \citenamefont
  {Weiss}}]{stellarstr}%
  \BibitemOpen
  \bibfield  {author} {\bibinfo {author} {\bibfnamefont {R}~\bibnamefont
  {Kippenhahn}}, \bibinfo {author} {\bibfnamefont {A.}~\bibnamefont {Weigert}},
  \ and\ \bibinfo {author} {\bibfnamefont {A.}~\bibnamefont {Weiss}},\ }\href
  {\doibase 10.1007/978-3-642-30304-3} {\emph {\bibinfo {title} {Stellar
  Structure and Evolution}}},\ \bibinfo {edition} {2nd}\ ed.\ (\bibinfo
  {publisher} {Springer-Verlag Berlin Heidelberg},\ \bibinfo {year}
  {2012})\BibitemShut {NoStop}%
\bibitem [{\citenamefont {Kepler}\ \emph {et~al.}(2007)\citenamefont {Kepler},
  \citenamefont {Kleinman}, \citenamefont {Nitta}, \citenamefont {Koester},
  \citenamefont {Castanheira}, \citenamefont {Giovannini}, \citenamefont
  {Costa},\ and\ \citenamefont {Althaus}}]{10.1111/j.1365-2966.2006.11388.x}%
  \BibitemOpen
  \bibfield  {author} {\bibinfo {author} {\bibfnamefont {S.~O.}\ \bibnamefont
  {Kepler}}, \bibinfo {author} {\bibfnamefont {S.~J.}\ \bibnamefont
  {Kleinman}}, \bibinfo {author} {\bibfnamefont {A.}~\bibnamefont {Nitta}},
  \bibinfo {author} {\bibfnamefont {D.}~\bibnamefont {Koester}}, \bibinfo
  {author} {\bibfnamefont {B.~G.}\ \bibnamefont {Castanheira}}, \bibinfo
  {author} {\bibfnamefont {O.}~\bibnamefont {Giovannini}}, \bibinfo {author}
  {\bibfnamefont {A.~F.~M.}\ \bibnamefont {Costa}}, \ and\ \bibinfo {author}
  {\bibfnamefont {L.}~\bibnamefont {Althaus}},\ }\bibfield  {title} {\enquote
  {\bibinfo {title} {{White dwarf mass distribution in the SDSS}},}\ }\href
  {\doibase 10.1111/j.1365-2966.2006.11388.x} {\bibfield  {journal} {\bibinfo
  {journal} {Monthly Notices of the Royal Astronomical Society}\ }\textbf
  {\bibinfo {volume} {375}},\ \bibinfo {pages} {1315--1324} (\bibinfo {year}
  {2007})},\ \Eprint
  {http://arxiv.org/abs/https://academic.oup.com/mnras/article-pdf/375/4/1315/18673650/mnras0375-1315.pdf}
  {https://academic.oup.com/mnras/article-pdf/375/4/1315/18673650/mnras0375-1315.pdf}
  \BibitemShut {NoStop}%
\bibitem [{\citenamefont {Tang}\ \emph {et~al.}(2014)\citenamefont {Tang},
  \citenamefont {Bildsten}, \citenamefont {Wolf}, \citenamefont {Li},
  \citenamefont {Kong}, \citenamefont {Cao}, \citenamefont {Cenko},
  \citenamefont {Cia}, \citenamefont {Kasliwal}, \citenamefont {Kulkarni},
  \citenamefont {Laher}, \citenamefont {Masci}, \citenamefont {Nugent},
  \citenamefont {Perley}, \citenamefont {Prince},\ and\ \citenamefont
  {Surace}}]{Tang_2014}%
  \BibitemOpen
  \bibfield  {author} {\bibinfo {author} {\bibfnamefont {Sumin}\ \bibnamefont
  {Tang}}, \bibinfo {author} {\bibfnamefont {Lars}\ \bibnamefont {Bildsten}},
  \bibinfo {author} {\bibfnamefont {William~M.}\ \bibnamefont {Wolf}}, \bibinfo
  {author} {\bibfnamefont {K.~L.}\ \bibnamefont {Li}}, \bibinfo {author}
  {\bibfnamefont {Albert K.~H.}\ \bibnamefont {Kong}}, \bibinfo {author}
  {\bibfnamefont {Yi}~\bibnamefont {Cao}}, \bibinfo {author} {\bibfnamefont
  {S.~Bradley}\ \bibnamefont {Cenko}}, \bibinfo {author} {\bibfnamefont
  {Annalisa~De}\ \bibnamefont {Cia}}, \bibinfo {author} {\bibfnamefont
  {Mansi~M.}\ \bibnamefont {Kasliwal}}, \bibinfo {author} {\bibfnamefont
  {Shrinivas~R.}\ \bibnamefont {Kulkarni}}, \bibinfo {author} {\bibfnamefont
  {Russ~R.}\ \bibnamefont {Laher}}, \bibinfo {author} {\bibfnamefont {Frank}\
  \bibnamefont {Masci}}, \bibinfo {author} {\bibfnamefont {Peter~E.}\
  \bibnamefont {Nugent}}, \bibinfo {author} {\bibfnamefont {Daniel~A.}\
  \bibnamefont {Perley}}, \bibinfo {author} {\bibfnamefont {Thomas~A.}\
  \bibnamefont {Prince}}, \ and\ \bibinfo {author} {\bibfnamefont {Jason}\
  \bibnamefont {Surace}},\ }\bibfield  {title} {\enquote {\bibinfo {title}
  {{AN} {ACCRETING} {WHITE} {DWARF} {NEAR} {THE} {CHANDRASEKHAR} {LIMIT} {IN}
  {THE} {ANDROMEDA} {GALAXY}},}\ }\href {\doibase 10.1088/0004-637x/786/1/61}
  {\bibfield  {journal} {\bibinfo  {journal} {The Astrophysical Journal}\
  }\textbf {\bibinfo {volume} {786}},\ \bibinfo {pages} {61} (\bibinfo {year}
  {2014})}\BibitemShut {NoStop}%
\end{thebibliography}%

\end{document}